\pdfoutput=1
\documentclass[longauth]{aa}
\usepackage{natbib}
\usepackage{url,hyperref}
\usepackage{footmisc}
\usepackage{wrapfig,lipsum,booktabs}
\usepackage{txfonts}
\makeatletter
\renewcommand*{\@fnsymbol}[1]{\ifcase#1\or*\or$\dagger$\or$\ddagger$\or**\or$\dagger\dagger$\or$\ddagger\ddagger$\fi}
\makeatother
\usepackage{siunitx}
\usepackage{amssymb,amsmath}
\usepackage{makecell}
\usepackage[switch]{lineno}
\usepackage{amsmath}

\hypersetup{
        bookmarks=true,         % show bookmarks bar?
    unicode=false,          % non-Latin characters in AcrobatÕs bookmarks
    pdftoolbar=true,        % show AcrobatÕs toolbar?
    pdfmenubar=true,        % show AcrobatÕs menu?
    pdffitwindow=false,     % window fit to page when opened
    pdfstartview={FitH},    % fits the width of the page to the window
   pdfnewwindow=true,      % links in new window
    colorlinks=true,        % false: boxed links; true: colored links
   linkcolor=blue,          % color of internal links --> blue
      citecolor=blue,        % color of links to bibliography --> blue
      urlcolor=blue,           % color of external links --> blue      
}
\usepackage{ulem}

\begin{document}

\title{TeV flaring activity of the AGN PKS\,0625$-$354 in November 2018}
%\date{Received date / Accepted date }
\date{\today}

\author{H.E.S.S. Collaboration
\and F.~Aharonian \inst{\ref{DIAS},\ref{MPIK},\ref{Yerevan}}
\and F.~Ait~Benkhali \inst{\ref{LSW}}
\and J.~Aschersleben \inst{\ref{Groningen}}
\and H.~Ashkar \inst{\ref{LLR}}
\and M.~Backes \inst{\ref{UNAM},\ref{NWU}}
\and A.~Baktash \inst{\ref{UHH}}
\and V.~Barbosa~Martins \inst{\ref{DESY}}
\and J.~Barnard \inst{\ref{UFS}}
\and R.~Batzofin \inst{\ref{UP}}
\and Y.~Becherini \inst{\ref{APC},\ref{Linnaeus}}
\and D.~Berge \inst{\ref{DESY},\ref{HUB}}
\and K.~Bernl\"ohr \inst{\ref{MPIK}}
\and B.~Bi \inst{\ref{IAAT}}
\and M.~B\"ottcher \inst{\ref{NWU}}
\and C.~Boisson \inst{\ref{LUTH}}
\and J.~Bolmont \inst{\ref{LPNHE}}
\and M.~de~Bony~de~Lavergne \inst{\ref{CEA}}
\and J.~Borowska \inst{\ref{HUB}}
\and F.~Bradascio \inst{\ref{CEA}}
\and M.~Breuhaus \inst{\ref{MPIK}}
\and R.~Brose \inst{\ref{DIAS}}
\and A.~Brown \inst{\ref{Oxford}}
\and F.~Brun \inst{\ref{CEA}}
\and B.~Bruno \inst{\ref{ECAP}}
\and T.~Bulik \inst{\ref{UWarsaw}}
\and C.~Burger-Scheidlin \inst{\ref{DIAS}}
\and T.~Bylund \inst{\ref{CEA}}
\and S.~Caroff \inst{\ref{LAPP}}
\and S.~Casanova \inst{\ref{IFJPAN}}
\and R.~Cecil \inst{\ref{UHH}}
\and J.~Celic \inst{\ref{ECAP}}
\and M.~Cerruti \inst{\ref{APC}}
\and T.~Chand \inst{\ref{NWU}}
\and S.~Chandra \inst{\ref{NWU}}
\and A.~Chen \inst{\ref{Wits}}
\and J.~Chibueze \inst{\ref{NWU}}
\and O.~Chibueze \inst{\ref{NWU}}
\and G.~Cotter \inst{\ref{Oxford}}
\and J.~Damascene~Mbarubucyeye \inst{\ref{DESY}}
\and I.D.~Davids \inst{\ref{UNAM}}
\and J.~Djuvsland \inst{\ref{MPIK}}
\and A.~Dmytriiev \inst{\ref{NWU}}
\and V.~Doroshenko \inst{\ref{IAAT}}
\and K.~Egberts \inst{\ref{UP}}
\and S.~Einecke \inst{\ref{Adelaide}}
\and J.-P.~Ernenwein \inst{\ref{CPPM}}
\and G.~Fontaine \inst{\ref{LLR}}
\and M.~F\"u{\ss}ling \inst{\ref{DESY}}
\and S.~Funk \inst{\ref{ECAP}}
\and S.~Gabici \inst{\ref{APC}}
\and S.~Ghafourizadeh \inst{\ref{LSW}}
\and G.~Giavitto \inst{\ref{DESY}}
\and D.~Glawion\protect\footnotemark[1] \inst{\ref{ECAP}}
\and J.F.~Glicenstein \inst{\ref{CEA}}
\and J.~Glombitza \inst{\ref{ECAP}}
\and P.~Goswami \inst{\ref{APC}}
\and G.~Grolleron \inst{\ref{LPNHE}}
\and L.~Haerer \inst{\ref{MPIK}}
\and J.A.~Hinton \inst{\ref{MPIK}}
\and T.~L.~Holch \inst{\ref{DESY}}
\and M.~Holler \inst{\ref{Innsbruck}}
\and D.~Horns \inst{\ref{UHH}}
\and M.~Jamrozy \inst{\ref{UJK}}
\and F.~Jankowsky \inst{\ref{LSW}}
\and V.~Joshi \inst{\ref{ECAP}}
\and I.~Jung-Richardt \inst{\ref{ECAP}}
\and E.~Kasai \inst{\ref{UNAM}}
\and K.~Katarzy{\'n}ski \inst{\ref{NCUT}}
\and R.~Khatoon \inst{\ref{NWU}}
\and B.~Kh\'elifi \inst{\ref{APC}}
\and W.~Klu\'{z}niak \inst{\ref{NCAC}}
\and Nu.~Komin \inst{\ref{Wits}}
\and K.~Kosack \inst{\ref{CEA}}
\and D.~Kostunin \inst{\ref{DESY}}
\and R.G.~Lang \inst{\ref{ECAP}}
\and S.~Le~Stum \inst{\ref{CPPM}}
\and F.~Leitl \inst{\ref{ECAP}}
\and A.~Lemi\`ere \inst{\ref{APC}}
\and J.-P.~Lenain \inst{\ref{LPNHE}}
\and F.~Leuschner \inst{\ref{IAAT}}
\and A.~Luashvili \inst{\ref{LUTH}}
\and J.~Mackey \inst{\ref{DIAS}}
\and R.~Marx \inst{\ref{LSW}}
\and A.~Mehta \inst{\ref{DESY}}
\and M.~Meyer \inst{\ref{UHH}}
\and A.~Mitchell \inst{\ref{ECAP}}
\and R.~Moderski \inst{\ref{NCAC}}
\and A.~Montanari \inst{\ref{LSW}}
\and E.~Moulin \inst{\ref{CEA}}
\and M.~de~Naurois \inst{\ref{LLR}}
\and J.~Niemiec \inst{\ref{IFJPAN}}
\and P.~O'Brien \inst{\ref{Leicester}}
\and S.~Ohm \inst{\ref{DESY}}
\and L.~Olivera-Nieto \inst{\ref{MPIK}}
\and E.~de~Ona~Wilhelmi \inst{\ref{DESY}}
\and M.~Ostrowski \inst{\ref{UJK}}
\and S.~Panny \inst{\ref{Innsbruck}}
\and R.D.~Parsons \inst{\ref{HUB}}
\and S.~Pita \inst{\ref{APC}}
\and D.A.~Prokhorov \inst{\ref{Amsterdam}}
\and G.~P\"uhlhofer \inst{\ref{IAAT}}
\and M.~Punch \inst{\ref{APC}}
\and A.~Quirrenbach \inst{\ref{LSW}}
\and P.~Reichherzer \inst{\ref{CEA}}
\and A.~Reimer \inst{\ref{Innsbruck}}
\and O.~Reimer \inst{\ref{Innsbruck}}
\and H.~Ren \inst{\ref{MPIK}}
\and F.~Rieger \inst{\ref{MPIK}}
\and B.~Rudak \inst{\ref{NCAC}}
\and V.~Sahakian \inst{\ref{Yerevan_Phys}}
\and H.~Salzmann \inst{\ref{IAAT}}
\and D.A.~Sanchez \inst{\ref{LAPP}}
\and M.~Sasaki \inst{\ref{ECAP}}
\and F.~Sch\"ussler \inst{\ref{CEA}}
\and H.M.~Schutte \inst{\ref{NWU}}
\and J.N.S.~Shapopi \inst{\ref{UNAM}}
\and H.~Sol \inst{\ref{LUTH}}
\and A.~Specovius \inst{\ref{ECAP}}
\and S.~Spencer \inst{\ref{ECAP}}
\and {\L.}~Stawarz \inst{\ref{UJK}}
\and R.~Steenkamp \inst{\ref{UNAM}}
\and S.~Steinmassl \inst{\ref{MPIK}}
\and K.~Streil \inst{\ref{ECAP}}
\and I.~Sushch \inst{\ref{NWU}}
\and H.~Suzuki \inst{\ref{Konan}}
\and T.~Takahashi \inst{\ref{KAVLI}}
\and T.~Tanaka \inst{\ref{Konan}}
\and C.~van~Eldik \inst{\ref{ECAP}}
\and M.~Vecchi \inst{\ref{Groningen}}
\and J.~Veh \inst{\ref{ECAP}}
\and C.~Venter \inst{\ref{NWU}}
\and S.J.~Wagner \inst{\ref{LSW}}
\and A.~Wierzcholska\protect\footnotemark[1]  \inst{\ref{IFJPAN}}
\and M.~Zacharias \inst{\ref{LSW},\ref{NWU}}
\and D.~Zargaryan \inst{\ref{DIAS}}
\and A.A.~Zdziarski \inst{\ref{NCAC}}
\and A.~Zech \inst{\ref{LUTH}}
\and S.~Zouari \inst{\ref{APC}}
\and N.~\.Zywucka \inst{\ref{NWU}}
}

\institute{
Dublin Institute for Advanced Studies, 31 Fitzwilliam Place, Dublin 2, Ireland \label{DIAS} \and
Max-Planck-Institut f\"ur Kernphysik, P.O. Box 103980, D 69029 Heidelberg, Germany \label{MPIK} \and
Yerevan State University,  1 Alek Manukyan St, Yerevan 0025, Armenia \label{Yerevan} \and
Landessternwarte, Universit\"at Heidelberg, K\"onigstuhl, D 69117 Heidelberg, Germany \label{LSW} \and
Kapteyn Astronomical Institute, University of Groningen, Landleven 12, 9747 AD Groningen, The Netherlands \label{Groningen} \and
Laboratoire Leprince-Ringuet, École Polytechnique, CNRS, Institut Polytechnique de Paris, F-91128 Palaiseau, France \label{LLR} \and
University of Namibia, Department of Physics, Private Bag 13301, Windhoek 10005, Namibia \label{UNAM} \and
Centre for Space Research, North-West University, Potchefstroom 2520, South Africa \label{NWU} \and
Universit\"at Hamburg, Institut f\"ur Experimentalphysik, Luruper Chaussee 149, D 22761 Hamburg, Germany \label{UHH} \and
Deutsches Elektronen-Synchrotron DESY, Platanenallee 6, 15738 Zeuthen, Germany \label{DESY} \and
Department of Physics, University of the Free State,  PO Box 339, Bloemfontein 9300, South Africa \label{UFS} \and
Institut f\"ur Physik und Astronomie, Universit\"at Potsdam,  Karl-Liebknecht-Strasse 24/25, D 14476 Potsdam, Germany \label{UP} \and
Université de Paris, CNRS, Astroparticule et Cosmologie, F-75013 Paris, France \label{APC} \and
Department of Physics and Electrical Engineering, Linnaeus University,  351 95 V\"axj\"o, Sweden \label{Linnaeus} \and
Institut f\"ur Physik, Humboldt-Universit\"at zu Berlin, Newtonstr. 15, D 12489 Berlin, Germany \label{HUB} \and
Institut f\"ur Astronomie und Astrophysik, Universit\"at T\"ubingen, Sand 1, D 72076 T\"ubingen, Germany \label{IAAT} \and
Laboratoire Univers et Théories, Observatoire de Paris, Université PSL, CNRS, Université Paris Cité, 5 Pl. Jules Janssen, 92190 Meudon, France \label{LUTH} \and
Sorbonne Universit\'e, Universit\'e Paris Diderot, Sorbonne Paris Cit\'e, CNRS/IN2P3, Laboratoire de Physique Nucl\'eaire et de Hautes Energies, LPNHE, 4 Place Jussieu, F-75252 Paris, France \label{LPNHE} \and
IRFU, CEA, Universit\'e Paris-Saclay, F-91191 Gif-sur-Yvette, France \label{CEA} \and
University of Oxford, Department of Physics, Denys Wilkinson Building, Keble Road, Oxford OX1 3RH, UK \label{Oxford} \and
Friedrich-Alexander-Universit\"at Erlangen-N\"urnberg, Erlangen Centre for Astroparticle Physics, Nikolaus-Fiebiger-Str. 2, 91058 Erlangen, Germany \label{ECAP} \and
Astronomical Observatory, The University of Warsaw, Al. Ujazdowskie 4, 00-478 Warsaw, Poland \label{UWarsaw} \and
Université Savoie Mont Blanc, CNRS, Laboratoire d'Annecy de Physique des Particules - IN2P3, 74000 Annecy, France \label{LAPP} \and
Instytut Fizyki J\c{a}drowej PAN, ul. Radzikowskiego 152, 31-342 Krak{\'o}w, Poland \label{IFJPAN} \and
School of Physics, University of the Witwatersrand, 1 Jan Smuts Avenue, Braamfontein, Johannesburg, 2050 South Africa \label{Wits} \and
School of Physical Sciences, University of Adelaide, Adelaide 5005, Australia \label{Adelaide} \and
Aix Marseille Universit\'e, CNRS/IN2P3, CPPM, Marseille, France \label{CPPM} \and
Universit\"at Innsbruck, Institut f\"ur Astro- und Teilchenphysik, Technikerstraße 25, 6020 Innsbruck, Austria \label{Innsbruck} \and
Obserwatorium Astronomiczne, Uniwersytet Jagiello{\'n}ski, ul. Orla 171, 30-244 Krak{\'o}w, Poland \label{UJK} \and
Institute of Astronomy, Faculty of Physics, Astronomy and Informatics, Nicolaus Copernicus University,  Grudziadzka 5, 87-100 Torun, Poland \label{NCUT} \and
Nicolaus Copernicus Astronomical Center, Polish Academy of Sciences, ul. Bartycka 18, 00-716 Warsaw, Poland \label{NCAC} \and
Department of Physics and Astronomy, The University of Leicester, University Road, Leicester, LE1 7RH, United Kingdom \label{Leicester} \and
GRAPPA, Anton Pannekoek Institute for Astronomy, University of Amsterdam,  Science Park 904, 1098 XH Amsterdam, The Netherlands \label{Amsterdam} \and
Yerevan Physics Institute, 2 Alikhanian Brothers St., 0036 Yerevan, Armenia \label{Yerevan_Phys} \and
Department of Physics, Konan University, 8-9-1 Okamoto, Higashinada, Kobe, Hyogo 658-8501, Japan \label{Konan} \and
Kavli Institute for the Physics and Mathematics of the Universe (WPI), The University of Tokyo Institutes for Advanced Study (UTIAS), The University of Tokyo, 5-1-5 Kashiwa-no-Ha, Kashiwa, Chiba, 277-8583, Japan \label{KAVLI}
}

\offprints{H.E.S.S.~collaboration,
\protect\\\email{\href{mailto:contact.hess@hess-experiment.eu}{contact.hess@hess-experiment.eu};}
\protect\\\protect\footnotemark[1] Corresponding author
}

\abstract{Most $\gamma$-ray detected active galactic nuclei  are blazars with one of their relativistic jets pointing towards the Earth. Only a few objects belong to the class of radio galaxies or misaligned blazars. 
   Here, we investigate the nature of the object PKS\,0625$-$354, its  $\gamma$-ray flux and spectral variability and its broad-band spectral emission with observations from H.E.S.S., Fermi-LAT, Swift-XRT, and UVOT taken in November 2018. 
   The H.E.S.S. light curve above 200\,GeV shows an outburst in the first night of observations followed by a declining flux with a  halving time scale of 5.9\,h. The $\gamma\gamma$-opacity constrains the upper limit of the angle between the jet and the line of sight to $\sim10^\circ$. The broad-band spectral energy distribution shows two humps and can be well fitted with a single-zone synchrotron self Compton emission model. 

  We conclude that PKS\,0625$-$354, as an object showing clear features of both blazars and radio galaxies, can be classified as an intermediate active galactic nuclei.
   Multi-wavelength studies of such intermediate objects exhibiting features of both blazars and radio galaxies are sparse but crucial for the understanding of the broad-band emission of $\gamma$-ray detected active galactic nuclei in general. 
     } 
 

   \authorrunning{First author et al.}
   \titlerunning{TeV flaring of the active galactic nucleus PKS\,0625$-$354}         
               
   \maketitle

\def\pks0625{PKS\,0625$-$354}
\def\hess{H.E.S.S.}

\section{Introduction}
An active galactic nucleus (AGN) is believed to host a supermassive black hole in its center which is surrounded by an accretion disc, fast and slow-moving clouds (corresponding to broad- and narrow-line regions), and a dusty torus. 
In the very high energy (VHE, $E>100$\,GeV) $\gamma$-ray range, about 80 active galactic nuclei have been detected so far\footnote{\url{http://tevcat.uchicago.edu/}}. 
In radio-loud AGNs, a pair of plasma jets extend perpendicular to the accretion disc. Most of these objects discovered in the $\gamma$-ray regime fall into the class of blazars.
They are characterized by  strong variability in all energy bands and from sub-day up to year-long time scales \citep[e.g.][]{WagnerW, 2155flare}.
According to the unified model of radio-loud AGN \citep{urry95}, blazars are  AGNs that are viewed under a small angle between the jet axis and the line of sight.
Hence, strong Doppler beaming is expected to play an important role in the explanation of the properties.
In contrast, radio galaxies are viewed under a larger angle to the jet axis. The Doppler factor, defined as $\delta=[\Gamma (1-\beta\cos\theta)]^{-1}$, where $\Gamma=(\sqrt{1-\beta^2})^{-1}$, is the Lorentz factor, $\beta$ the velocity of the jet in units of the speed of light and $\theta$ the jet viewing angle,
is smaller compared to blazars.\\
Currently, key topics in AGN research are the connection between the jet and the black hole, the jet base, the acceleration and radiation physics in the jet, and the origin and location of the high-energy emission. Measuring the emission of AGN up to the VHE regime is essential for investigations of the underlying physical processes. 
Among extragalactic TeV  $\gamma$-ray emitting sources, only a few belong to the class of radio galaxies. These are  Centaurus A \citep{aharonian09}, M87 \citep{aharonian03}, NGC\,1275 \citep{aleksic12}, and 3C\,264 \citep{Archer_264}. The small number of these objects can be understood in terms of the Doppler boosting effect. The emission of the blazars gets strongly boosted by a high Doppler factor so that the flux and the energy of the $\gamma$ rays get amplified leading to a higher detection probability. For misaligned jets, due to a larger angle, this amplification is only moderate. Only a few misaligned blazars have been detected. The active galaxies IC\,310 \citep{aleksic10} and PKS\,0625$-$354 \citep{PKS0625discovery},  are  showing characteristics of both AGN types.    \\

\pks0625 is a Fanaroff-Riley \citep[FR,][]{Fanaroff} type\,I radio galaxy \citep{ojha10}, located at a redshift of $z=0.056$ \citep{Lauer14}, and its host galaxy is classified as a  low-ionization nuclear emission-line region \citep[LINER][]{lewis03}.
While the optical spectrum is that  of a BL\,Lac object \citep{wills04}, the kpc radio morphology clearly shows two extended lobe structures typical for a  radio galaxy \citep{angioni19}.
Radio observations revealed superluminal motion  with $v_{app}\approx(2.9\pm0.9)\,c$ \citep{angioni19}. The jet-to-counterjet ratio of the pc-scale radio jet  limits the viewing angle to $\theta<53^\circ$, which includes the possibility of an intermediate jet orientation between the radio galaxy and blazar classes \citep{angioni19}.

H.E.S.S. reported on the detection of \pks0625 in the VHE $\gamma$-ray band based on 5.5\,hrs  of good-quality data collected in 2012 \citep{PKS0625discovery}. The observed TeV gamma-ray photon spectral index of the object is consistent with a simple power-law index of $2.8\pm0.5$ (without correction for absorption due to the extragalactic background light). The flux measured by H.E.S.S. above 580\,GeV is $\sim 4\%$ of the Crab Nebula flux.  No VHE $\gamma$-ray variability was found by \cite{PKS0625discovery}.
Taking into account the redshift of the galaxy, it is among those with the brightest absolute luminosities from the VHE-detected non-blazar objects. 

In this paper we present the results from H.E.S.S. observations of \pks0625 as part of the first ever dedicated multi-wavelength campaign including \textit{Swift}-UVOT (optical/UV), \textit{Swift}-XRT (X-ray), and \textit{Fermi-LAT} ($\gamma$-ray) measurements conducted in November 2018 (in Sect. 2). In Sect. 3, we discuss the flux variability in the VHE $\gamma$-ray band and the simultaneous broad-band spectral energy distribution.

%--------------------------------------------------------------------
\section{Observations and results}

\subsection{VHE $\gamma$-ray observations and analysis}

H.E.S.S. (High Energy Stereoscopic System) is an array of five Imaging Air Cherenkov Telescopes located in the Khomas Highland of Namibia at an altitude of about 1800\,m a.s.l. Four telescopes with 12\,m in diameter are arranged in a square with a side length of 120\,m, and a larger  28\,m mirror dish diameter is placed in the center of the square (CT5 hereafter). Each telescope measures the Cherenkov light produced in air showers caused by the interaction of $\gamma$ rays or cosmic rays in the atmosphere of the Earth. Images of the showers are used to determine if they originate from $\gamma$-rays or hadrons, as well as to reconstruct the arrival direction and the energy. 

In November 2018, \pks0625 was observed by the full array in which events are triggered by CT5 and at least two of the smaller telescopes. The observation started on  November 1, 2018 (MJD 58424.01) as part of a program to search for the TeV flux variability of \pks0625. The real-time analysis running in parallel to the observation \citep{balzer14} indicated a significant detection of the target ($\sim 5\sigma$). Therefore, further observations were scheduled and performed in the following nights until MJD\,58433 resulting in a total of 17.5\,h of live time. The wobble mode was used for the observation where the source position is offset by 0.7$^\circ$ in right ascension or declination from the camera center. This allows simultaneous measurement of the background \citep{berge07}.   

The data were analyzed using a template-based analysis technique \citep{parsons14,parsons15}. The background produced by cosmic rays is rejected using a neural network-based scheme. The residual background contamination level of the source region is estimated with the ring background method for the maps and the reflected background method for the reconstruction of the spectrum and light curve \citep{berge07}. The background normalization factor is denoted with $\alpha_{\mathrm{Exp}}=\mathrm{Exposure}_\mathrm{ON}/\mathrm{Exposure}_\mathrm{OFF}$ with ON and OFF referring to the signal and the background, respectively. Differential and integral upper limits are derived following \citet{rolke} at the 95\% confidence level.   An independent analysis chain \citep{deNaurois2009} was used for the calibration and reconstruction as a crosscheck, which yielded compatible results. 

The ring background method results in the detection of 771 ON and 16881 OFF  events  ($\alpha_{\mathrm{Exp}}=0.0326$) yielding a significance of $8.7\sigma$. The reflected background method resulted in 771 ON and 12701 OFF events ($\alpha_{\mathrm{Exp}}= 0.0435$), and  $8.6\sigma$ of significance.

The reported results of the flux measurements, spectra, and upper limits include only statistical errors. For the systematic errors, we conservatively estimated an uncertainty of 20\% on the flux normalization and 0.1 on the photon spectral index \citep{aharonian06b}.

  \begin{figure}
   \centering
   \includegraphics[width=9.cm]{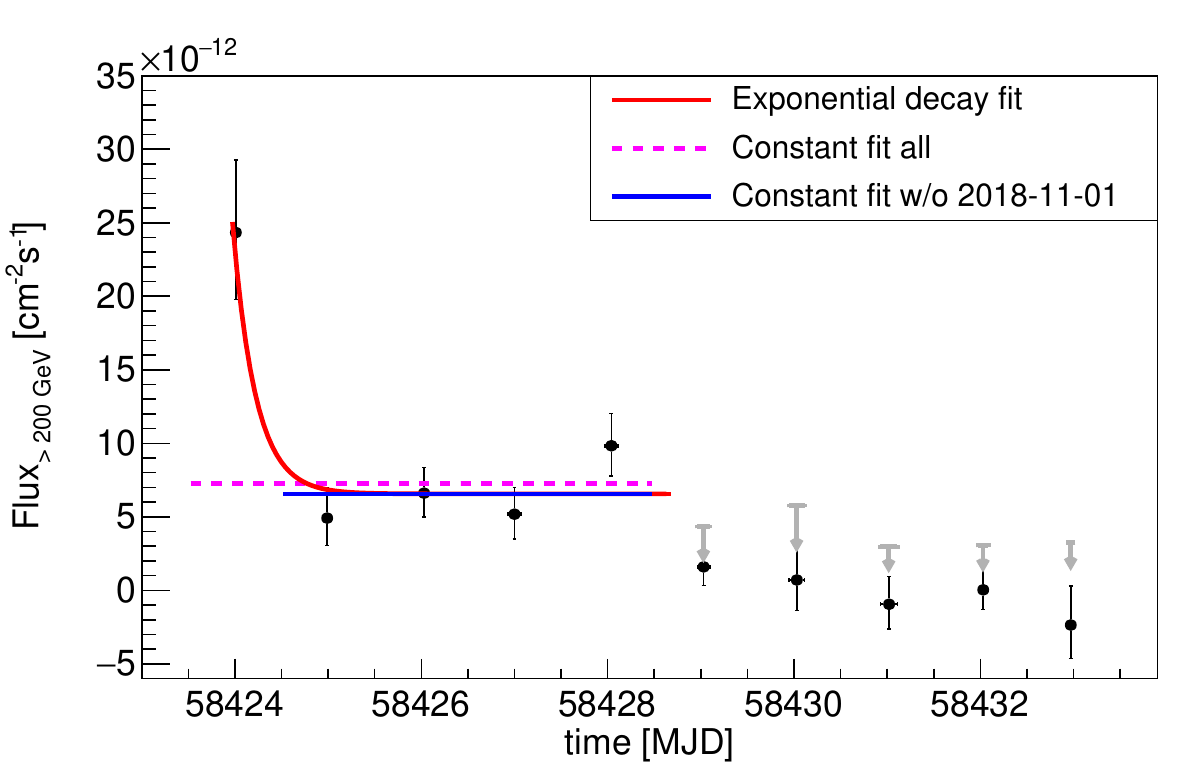}
   \caption{Nightly-binned light curve above 200\,GeV measured with the H.E.S.S. telescopes in November 2018 (black data points). Integrated upper limits (gray arrows) are given at a 95\% confidence level. The red solid line shows the exponential fit to the measured flux points excluding upper limits, while the pink dashed line shows a constant fit to those flux points, and the blue solid line is the constant fit to those points except the first night.}
              \label{VHELightcurve}%
    \end{figure}
    
 \begin{table}
\small
\caption{H.E.S.S. $\gamma$-ray flux measurements above 200\,GeV  from individual observation nights. The dates given in the first column correspond to the day at the beginning of the observation night. Column three indicates the total live time of the measurement during the night. Flux measurements and upper limits (UL) computed with a 95\% confidence level are given in columns four and five, respectively. 
}        
\label{table:1}     
\centering                        
\begin{tabular}{c c c c c}       
\hline\hline 
   date             &MJD    & $t$ & $F_{\mathrm{E}>0.2\,\mathrm{TeV}}$ & $F_{\mathrm{UL}, \mathrm{E}>0.2\,\mathrm{TeV}}$\\    
  & & [h] & [$10^{-12}$\,cm$^{-2}$\,s$^{-1}$] & [$10^{-12}$\,cm$^{-2}$\,s$^{-1}$] \\
\hline 
2018-11-01 & 58424.01 & 0.77 & $\,24.3_{-4.6}^{+4.9}$ & \\
2018-11-02 & 58424.99 & 1.75 & $\,\,\,\,4.9_{-1.9}^{+2.0}$ & \\
2018-11-03 & 58426.03 & 2.38 & $\,\,\,\,6.6_{-1.6}^{+1.7}$  &\\
2018-11-04 & 58427.00 & 2.61 & $\,\,\,\,5.2_{-1.7}^{+1.8}$  & \\
2018-11-05 & 58428.04 & 1.83 & $\,\,\,\,9.8_{-2.0}^{+2.2}$  & \\
2018-11-06 & 58429.03 & 3.03 & $\,\,\,\,1.6_{-1.3}^{+1.3}$  &$<4.3$ \\
2018-11-07 & 58430.03 & 0.91 & $\,\,\,\,0.7_{-2.1}^{+2.4}$  &$<5.8$\\
2018-11-08 & 58431.02 & 1.53 & $-0.9_{-1.7}^{+1.9}$  &$<3.0$\\
2018-11-09 & 58432.03 & 2.29 & $\,\,\,\,0.0_{-1.3}^{+1.5}$  &$<3.1$\\
2018-11-10 & 58432.97 & 0.92 & $-2.4_{-2.3}^{+2.7}$   &$<3.3$\\
\hline                                   
\end{tabular} 
\end{table}   

The light curve of the nightly integrated flux above 200\,GeV measured between 
2018 November 1 (MJD 58423.99) and 2018 November 10 (MJD 58432.99)
is shown in Fig.~\ref{VHELightcurve}. It was calculated assuming a photon spectral index of $\Gamma=3.0$ ($\mathrm{d}N/\mathrm{d}E\propto E^{-\Gamma}$) similar to the value found in \citet{PKS0625discovery}. 
Fluxes as well as integral upper limits of individual nights are given in Table~\ref{table:1}. 

We note here that PKS\,0625-354 was also observed with H.E.S.S. in 2019. However, 6.5\,hours of good-quality data collected during this period of MJD58579 and MJD58782-58793, resulted in no detection of the source, yielding only a flux upper limit  I($>$200\,GeV) $< 5\times10^{-12}$\,cm$^{-2}$\,s$^{-1}$.

In this section, we investigate the flux variability in the nightly-binned light curve. The study with smaller time bins showed a constant flux within one observation night for all nights. As no H.E.S.S. observations of \pks0625 were performed in the weeks before 2018-11-01,
we cannot make a statement about the start or the duration of the flare.

Fitting all  points with significance $>2\sigma$ in the light curve with a constant function yields a $\chi^{2}$ of 18.4 for 4 degrees of freedom (d.o.f.) corresponding to a $\chi^{2}$-probability of $1\times10^{-3}$. Excluding the data from the first observation night (2018-11-01) improves the fit probability to $0.3$.

In order to quantify the variability we estimate the time scale in which the flux changed, by fitting the light curve  with an exponential decay function \citep{Rani13, Baghmanyan17, Gasparyan18}:

\begin{equation}
F=F_{\mathrm{b}}+F_1\cdot e^{-|t-t_1|/\tau}.   
\end{equation}
Here, $F_{\mathrm{b}}$ is the baseline flux, $F_1$ the normalization flux at the time $t_1$, and $\tau$ is the flux variability time-scale. Fitting the nightly-binned flux points excluding the nights with low ($<2\sigma$) significance with Eq.~1 yields a flux variability time-scale of $\tau=(5.9\pm2.7)$\,h and $F_\mathrm{b}=(6.6\pm0.9)\times10^{-12}\,\mathrm{cm}^{-2}\mathrm{s}^{-1}$ with a $\chi^{2}$/d.o.f. of 4.2/3 corresponding to a $\chi^{2}$-probability of $0.2$. The fit is shown in Fig.~\ref{VHELightcurve}.
A different approach is used following, for example, \citet{Foschini2013,Brown2013,Coogan16},

\begin{equation}
\frac{F(t_1)}{F(t_0)}=2^{-|t-t_1|/\tau_\mathrm{H}},   
\end{equation}
where $F(t_1)$ and $F(t_0)$ are two flux measurements of  consecutive nights $t_0$ $t_1$ and $\tau_\mathrm{H}$ indicates the flux halving time-scale. The most significant flux change of $3.5\,\sigma$ is found between the first and the second observation night resulting in a flux halving time-scale of $\tau=(10.3\pm6.9)$\,h. We  note  that there is no clear indication of exponential decay in the data. Also, we do not know the starting point of the high flux state. For the rest of this work, we use $\tau=(5.9\pm2.7)$\,h as variability time-scale even though it can be only considered as an upper limit.

\begin{table*}

\caption{Results of power-law fit of the spectra measured with H.E.S.S. For  comparison, we give the converted flux at 0.40\,TeV in brackets for the 2012 measurement. }            
\label{table:VHESpecta}      
\centering                          
\begin{tabular}{c c c c c}        
\hline\hline
state    &$f_{0}\pm f_{\mathrm{stat}}\pm f_{\mathrm{syst}}$ & $E_0$ &$\Gamma_\mathrm{VHE}\pm\Gamma_{\mathrm{stat}}\pm\Gamma_{\mathrm{syst}}$ & energy range\\
         &$\times10^{-12}[\mathrm{TeV}^{-1}\,\mathrm{cm}^{-2}\,\mathrm{s}^{-1}]$& [TeV] &  & [TeV]\\
\hline
average  &\,\,\,$7.51\pm1.02\pm1.50$  & 0.40 & $2.83\pm0.26\pm0.10$ &  0.15 - 1.20\\  
high &$23.82\pm4.95\pm4.76$ & 0.44 &$2.90\pm0.49\pm0.10$ & 0.24 - 1.20\\  
low      &\,\,\,$6.27\pm1.02\pm1.25$   & 0.40 &$2.90\pm0.31\pm0.10$ & 0.15 - 1.20\\  
\hline
2012     &\,\,\,$0.58\pm0.22\pm0.12$   & 1.00 &$2.84\pm0.50\pm0.10$ &  0.2 - 10.0 \\
        & \,\,\,$(7.83\pm2.97\pm1.62)$       & (0.40) &                     &               \\
\hline      
\end{tabular}

\end{table*}

\begin{figure}
   \centering
   \includegraphics[width=9.cm]{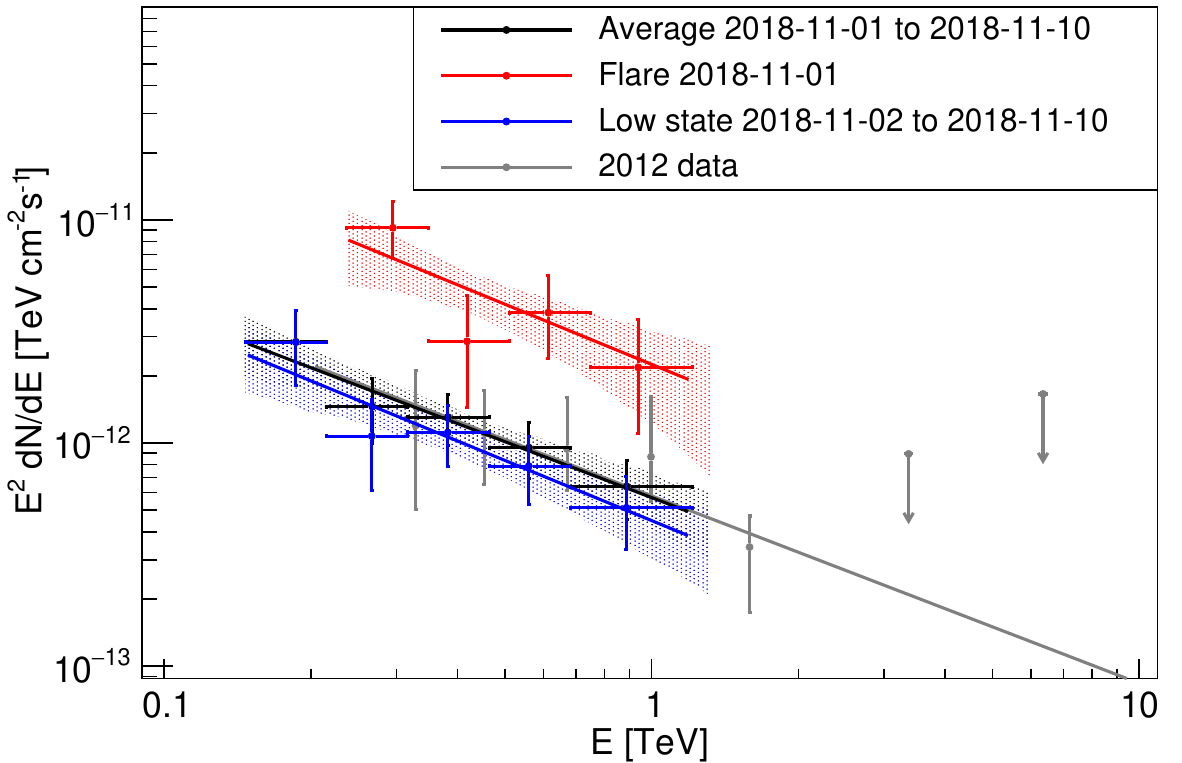}
   \caption{Spectral energy distributions measured with the H.E.S.S. telescopes in 2018 and 2012. The averaged spectrum is shown with black points /butterfly and a black fit line. The red data points, butterfly and the red line indicate the spectrum measured on 2018-11-01, while the data points /butterfly and the line in blue shows the spectrum computed from the remaining data taken in November 2018.  For comparison, we show the results obtained in 2012 in gray \citep{PKS0625discovery}. }
              \label{VHESEDs}%
    \end{figure}

Spectral energy distributions are shown in Fig.~\ref{VHESEDs} and results of the spectral fits are given in Table~\ref{table:VHESpecta}. They are described by a power-law function:

\begin{equation}
 \mathrm{d}N/\mathrm{d}E=f_0\cdot \left( \frac{E}{E_0} \right)^{-\Gamma_\mathrm{VHE}}, \end{equation}
where $f_0$ is the normalization at the energy $E_0$ and $\Gamma_\mathrm{VHE}$ is the photon spectral index. We calculate an averaged spectrum of all data in order to compare it with the results obtained from the 2012 measurements \citep{PKS0625discovery}. Furthermore, we compute a spectrum of the data from the night with the highest flux (2018-11-01) and one for the remaining data to study possible spectral index variations.

\subsection{HE $\gamma$-ray observations and analysis}\label{lat}
High energy $\gamma$-ray observations of PKS\,0625$-$354 (4FGL\,J0627.0$-$3529) were performed with the \textit{Fermi}-LAT detector. 
The data collected in the energy range from 100\,MeV to 500\,GeV were analyzed using standard ScienceTools with \verb|P8R3_SOURCE_V3| instrument response function. 
For the analysis, events within a 10$^{\circ}$ region of interest (ROI) centered on the source were selected. 
The binned maximum-likelihood method \citep{Mattox96} was applied in the analysis, with the Galactic diffuse background modeled using the gll$\_$iem$\_$v06 map cube.
All sources from the Fermi-LAT Third Source Catalog \citep[4FGL,][]{4FGL} inside the ROI of PKS\,0625$-$354 were modeled.

\begin{figure}
   \centering
   \includegraphics[width=8.5cm]{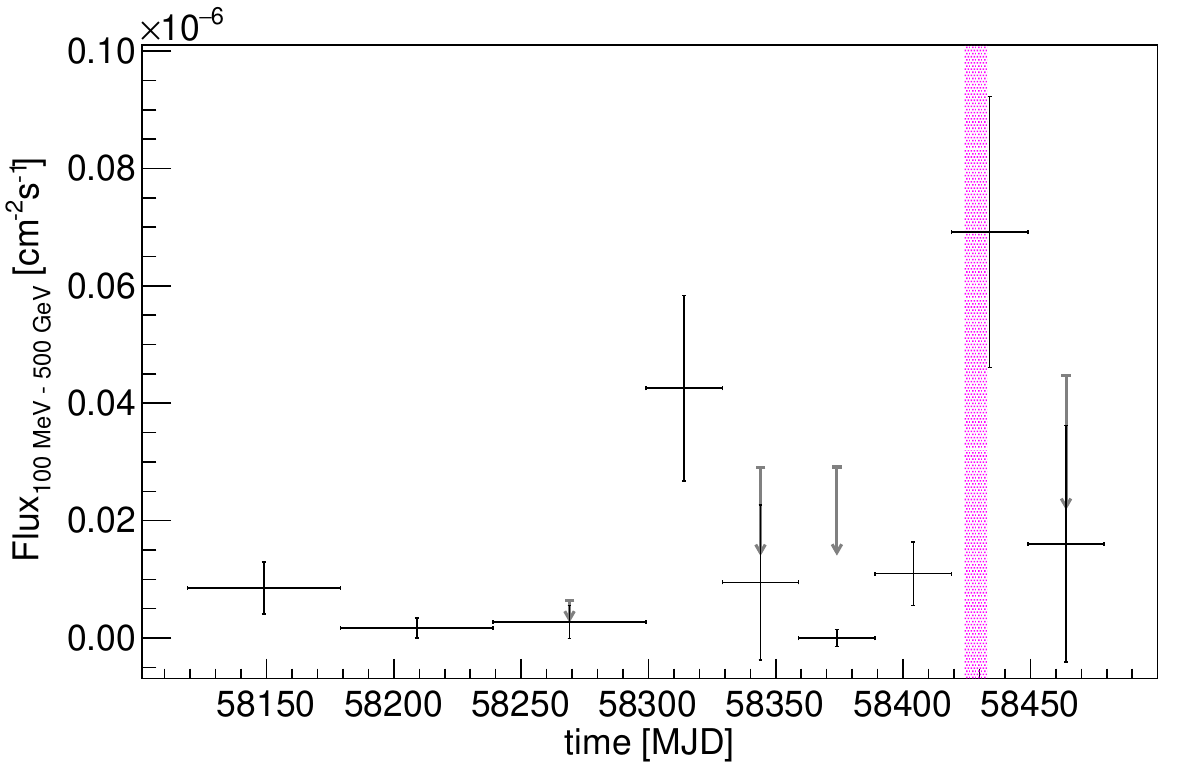}
   \caption{Light curve and flux upper limits measured with \textit{Fermi}-LAT in 2018 between 100\,MeV and 500\,GeV. A two-month binning is chosen for the first half of the year 2018 and a monthly binning for the second half of the year. The magenta area shows the time when H.E.S.S. observations took place. }
              \label{Fermi_LC}
              \end{figure}

Two time intervals were selected for the spectral analysis:  one overlapping with  the H.E.S.S. flaring state (MJD58421--58425), and one representing the H.E.S.S. low state period (MJD58424--58433). 
The flaring state mentioned here includes data centered on the flux maximum measured with H.E.S.S. and observations performed just before and after the H.E.S.S. flare.
The low state corresponds to the period after the H.E.S.S. flare and includes all of the consecutive VHE observations performed just after the flare. 
The two intervals selected for the Fermi-LAT spectra have a small overlap. Such a selection of the low-state \textit{Fermi}-LAT spectrum allows us, however, to have exactly the same periods covered by the VHE and HE $\gamma$-ray observations and cover the simultaneous X-ray observations as well. 
The spectral reconstruction of the Fermi-LAT low state is not affected if the time selection is shifted by 24\,hours to avoid overlap with the high state.

 In the case of the flaring state, the analysis performed resulted in a test statistic (TS) of 11.2, while for the low state a TS\,$=9.5$ is found.  
The flaring and low state are both well described by power-law spectra with a photon index of 2.15$\pm$0.36 and 2.08$\pm$0.31, and a normalization of (1.44$\pm$0.93) $\times$ 10$^{-12}$ MeV\,s$^{-1}\,\mathrm{cm}^{-2}$ and (0.62$\pm$0.35) $\times$ 10$^{-12}$\,MeV\,s$^{-1}\,\mathrm{cm}^{-2}$, respectively.

A corresponding butterfly is calculated for both cases and included in the broad-band spectral energy  distribution presented in Fig.\,\ref{SED_0625}.  A more detailed (binned) spectral and light curve analysis from of November 2018 is not reliable with the limited statistics. Furthermore, a statement about  potential spectral variation from the long-term spectrum is also inconclusive. Within the uncertainties the spectral indices agree with the long-term measurement. A long-term light curve analysis over the entire year of 2018 is shown in Fig~\ref{Fermi_LC}. A constant fit to the light curve yields a $\chi^{2}$/d.o.f. of 31.3/8 corresponding to a probability of $1.2\times10^{-4}$. Interestingly, the highest flux point is found in November 2018.

\subsection{X-ray observations and analysis}
{\it Swift}-XRT observations of PKS\,0625$-$354 in 2018 were carried out on November 3, 4, and 5,  for 1988\,s, 2149\,s, and 	1895\,s, respectively. 
All observations, corresponding to the ObsIDs of 00088133003--00088133005, were performed in  the photon counting mode (PC) in the energy range of 0.3–10\,keV. 

The data analysis was performed using the HEASOFT software  (version 6.29). 
Data were recalibrated using the standard \verb|xrtpipeline| procedure.
 For the spectral fitting \verb|xspec| was  used  \citep{Arnaud}.

All  data  were  binned to  have  at  least  30  counts  per  bin. Energy fluxes have  been  derived  by  fitting  each  single  observation  with a single power-law model with  a  Galactic  absorption  value  of {\it N}$_H$= 6.5$\cdot$10$^{20}$\,cm$^{-2}$ \citep[LAB survey,][]{Kalberla05}. 
 Fit parameters of  the spectral analysis of the X-ray observations are included in Table\,\ref{table_LC_swift}. The butterfly of the spectral analysis of the total data set is shown in Fig.~\ref{SED_0625}. 

\subsection{Optical and ultraviolet observations and analysis}
 Simultaneous optical and ultraviolet observations in six bands were performed with \textit{Swift}-UVOT: V (544\,nm), B (439\,nm), U (345\,nm), UVW1 (251\,nm), UVM2 (217\,nm) and UVW2 (188\,nm). 
 In the case of each of these observations, for the analysis, all photons in a circular region with a radius of 5\,arcsec  were taken into account. The background was determined also from a circular region located near the source region but in an area not contaminated by signals from nearby sources.
The instrumental magnitudes and the corresponding fluxes were calculated with the \verb|uvotsource| task.
The conversion into flux units was done using the factors from \cite{Poole08}.
The measured fluxes were corrected for the dust absorption using $E(B-V) = 0.0562$\,mag \citep{Schlafly11}  and  $A_{\lambda}/E(B-V)$, as provided by \cite{Giommi06}.
 All \textit{Swift}-UVOT measurements are listed in Table\,\ref{table:UVOT_2} and shown in Fig.~\ref{SED_0625}.

\begin{table*}
\caption{Swift-XRT X-ray spectral measurements from observations in  November 2018. Energy fluxes in the range 2-10\,keV were determined by a simple absorbed power-law fit. The photon spectral index $\Gamma$ is defined following $F=F_N/1\mathrm{keV}\times E^{-\Gamma}$. The absorption was kept fixed to the Galactic absorption with an equivalent column of hydrogen of $N_\mathrm{H}=6.5\times10^{20}$\,cm$^{2}$ as provided by \cite{Kalberla05}. }             
\label{table_LC_swift}     
\centering       

\begin{tabular}{c c c c c c c}       
\hline\hline
   Obs. ID  &MJD start     & Exps.  & $F_{2-10\,\mathrm{keV}}\times10^{-12}$ & $F_N\times10^{-3}$ & $\Gamma$  & $\chi^2$/d.o.f.\\    
		  	   & 	    & [s]    & [erg\,s$^{-1}$\,cm$^{-2}$] & [ph\,cm$^{-2}$\,s$^{-1}$\,keV$^{-1}$]  &  & \\
\hline 
00088133003 &  58425.02   & 1988.28  & $5.60\pm0.28$ & $1.93\pm0.10$ &$1.915\pm0.076$ &  1.01               \\
00088133004 &  58426.09	  &2149.43  & $4.94\pm0.23$  & $2.15\pm0.10$ & $2.072\pm0.071$  & 0.99\\
00088133005 & 58427.6     & 1895.63 & $5.04\pm0.24$  & $2.11\pm0.11$ & $2.046\pm0.069$  & 1.084\\
\hline
total       &              & 6033.34 &$5.03\pm0.20$   & $2.20\pm0.06$ & $2.07\pm0.04$  & 0.958 \\
\hline 
\end{tabular} 
\end{table*}

\begin{table*}
\caption{\textit{Swift}-UVOT Fluxes for different observations, corrected for the Galactic extinction. The columns present: (1) the observation ID number and (2)–(7) the observed fluxes  (in 10$^{-11}$\,erg\,cm$^{-2}$\,s$^{-1}$)} in V, B, U, UVW1, UVM2 and UVW2 bands, respectively.             
\label{table:UVOT_2}      
\centering                         
\begin{tabular}{c c c c c c c}        
\hline\hline
Observation ID & V & B & U & UVW1 & UVM2 & UVW2 \\
(1) & (2) & (3) & (4) & (5) &  (6) & (7)   \\
\hline     
00088133003  & 1.17$\pm$0.08 & 0.65$\pm$0.04 & 0.43$\pm$0.03 & 0.33$\pm$0.03 &  0.40$\pm$0.02 & 0.45$\pm$0.03   \\

00088133004  & 1.11$\pm$0.11 & 0.71$\pm$0.06 & 0.38$\pm$0.04 & 0.42$\pm$0.04 &  0.45$\pm$0.03& 0.45$\pm$0.03    \\

00088133005  & 1.16$\pm$0.11 & 0.69$\pm$0.07 & 0.39$\pm$0.05 & 0.38$\pm$0.03 &  0.43$\pm$0.03& 0.47$\pm$0.03    \\

\hline
\end{tabular}
\end{table*}

\section{Discussion}

\subsection{Implications from fast VHE $\gamma$-ray variability }

For this discussion we assume that the VHE emission originates from a spherical region with a radius $R$ at a redshift $z$. Due to causality argument, $R$ can be constrained by the variability time scale $\tau$ following

\begin{equation}
    R \leq c\tau\delta(1+z)^{-1},
\end{equation}
where $\delta$ is the Doppler factor and $c$ is the speed of light. For $\tau=(5.9\pm2.7)$\,h and $z=0.056$, this yields $R\leq\delta\times6.0\cdot10^{14}$\,cm. 
The mass of the central supermassive black hole in \pks0625 is estimated to be $10^{9.19\pm0.37}\,M_\odot$ \citep{bettoni03}, corresponding to $\sim2.1$\,h for the event horizon light crossing time\footnote{We calculated the event horizon light crossing time assuming a maximally rotating black hole (Kerr metric), thus $GM_\mathrm{BH}/c^3$.}, 
which is smaller than the observed variability time scale.

For a small emission region as estimated from the fast variability time scale, it is necessary to investigate if possible internal absorption via $\gamma\gamma$-pair production takes place. For this we estimate the optical depth for pair production using Eq.\,9 in \citet{abdo11} based on \citet{Dondi1995}:

\begin{equation}
  \tau_{\gamma\gamma}\sim\frac{\sigma_\mathrm{T}D_\mathrm{L}^2F_0\epsilon_\gamma(1+z)}{10Rm_e^2c^5\delta^5},
\end{equation}
where $\sigma_\mathrm{T}$ is the Thomson cross section, $m_e$ is the electron mass, and $D_\mathrm{L}$ is the luminosity distance.  For 1\,TeV photons ($\epsilon_\gamma=1$\,TeV in this case) and therefore, to significantly reduce pair production, $\tau_{\gamma\gamma}$ needs to be $<1$ (for $\tau_{\gamma\gamma}=1$ the flux is reduced by $1/e$). Taking the observed (though not simultaneous and not corrected for host galaxy emission) value from the 2MASS survey in the H-band  $F_\mathrm{\mathrm{2MASS, H}}= (1.67\pm0.13)\times10^{-11}$\,erg\,cm$^{-2}$\,s$^{-1}$ \citep{2MASS} for $F_0$ where the absorption of the TeV photons is supposed to happen, this yields a constraint for the Doppler factor of $\delta>6.7\pm0.5)$. \footnote{Reducing the assumed H-band flux down to 10\% of its value as suggested by the SSC modeling in Sec.~3.2 (approximately the host-galaxy corrected value), results in a Doppler factor of
 $\delta>(4.6\pm0.7)$.} Notably, the value of $\delta$ is robust to uncertainties in $R$ and $F_0$ as according to Eq.~(5) $\delta$ only depends weakly on these parameters, $\delta\propto(F_0/R)^{(1/5)}$.

\begin{figure}
   \centering
   \includegraphics[width=7.cm]{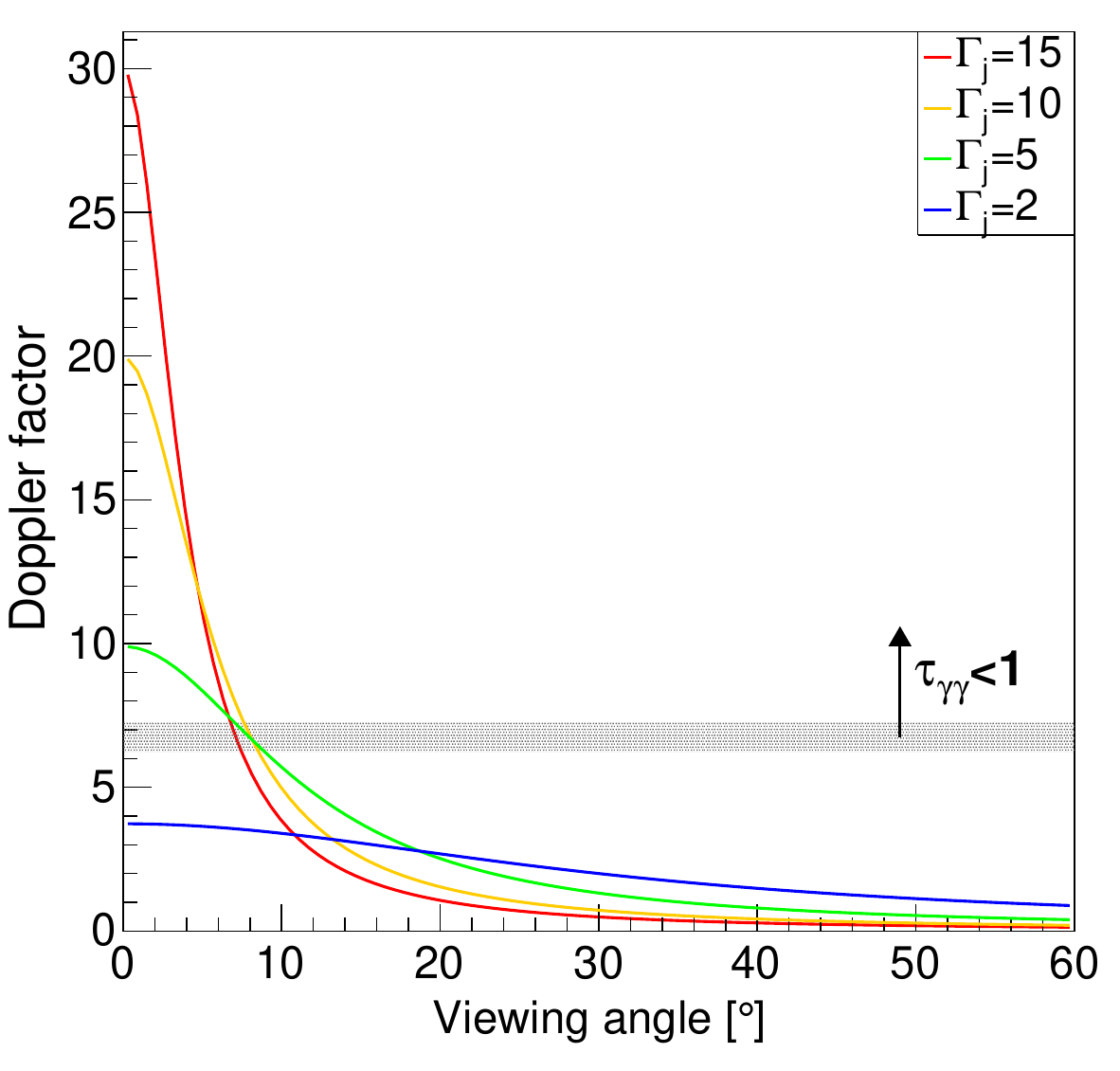}
   \caption{ Doppler factor as function of viewing angle plotted for different bulk Lorentz factors (red $\Gamma_\mathrm{j}=15$, yellow $\Gamma_\mathrm{j}=10$, green $\Gamma_\mathrm{j}=5$, and  blue $\Gamma_\mathrm{j}=2$ ). The ranges of Doppler factors for which the optical depth for $\gamma\gamma$-pair production constrained by the variability time scale is smaller than 1 is indicated with the black box and the arrow.}  
              \label{Doppler}%
    \end{figure}

Assuming that the $\gamma$-ray emission originates from a single zone, the constraint for the Doppler factor implies that viewing angles larger than $\sim10^\circ$ can be excluded for \pks0625 for any values of the bulk Lorentz factor (see Fig.~\ref{Doppler}). This is a stronger constraint than the limit obtained from very-long baseline interferometry observations in the radio band \citep{angioni19}.  Furthermore, if the viewing angle is not  small ($\theta>5^\circ$) as suggested by the large scale radio morphology, then a   small Lorentz factor (for example $\Gamma_\mathrm{j}=5$) provides a higher Doppler boosting. Instead, assuming a larger bulk Lorentz factor for the same viewing angle already results in a smaller Doppler boosting.

\subsection{Broad-band spectral energy distribution}

\begin{figure*}
   \centering
   \includegraphics[width=14.cm]{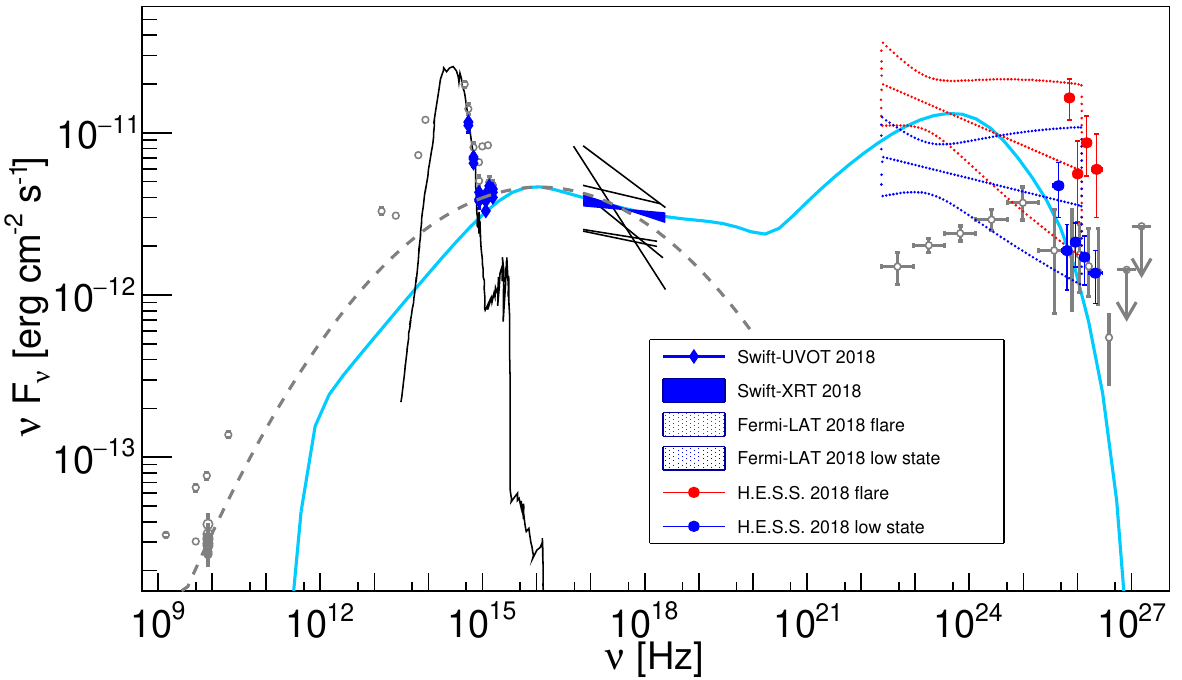}
   \caption{Broad-band spectral energy distribution of \pks0625 including (quasi-) simultaneous data (colored) as well as archival measurements shown with gray data points.   The HESS data points were corrected for EBL absorption using the model by \citet{dominguez11}. The red data points indicate the TeV high state on 2018-11-01.  The blue data points were obtained during a TeV low state after  2018-11-01. The red and blue unfilled butterflies are results from \textit{Fermi}-LAT measurements as described in Sect. \ref{lat}.
   The light blue line shows the model obtained with a single-zone SSC code from \citet{krawczynski04}. The dashed gray line is a log-parabola fit to Tanami data, the high frequency UVOT and the XRT data in order to identify the location of the first SED maximum and its amplitude. Archival measurements from \cite{PKS0625discovery},  best fits of historical  X-ray measurements \citep[gray lines, as provided by][]{PKS0625discovery}, and \cite{angioni19} are included. For the historical X-ray results, we show only the best fit functions.  The black line indicates the host galaxy flux model as presented in \cite{PKS0625discovery}.}
              \label{SED_0625}%
    \end{figure*}

The spectral energy distribution (SED) of a radio-loud AGN can be  explained by non-thermal emission of particles accelerated at shock waves in the jet  resulting from plasma density fluctuations  \citep{blandford79}. Generally, two peaks are measured. The low energy part of the SED is explained by synchrotron radiation of electrons and positrons. The origin of the high energy hump is widely discussed in the literature. It has been modeled with leptonic processes, i.e., inverse-Compton scattering 
\citep{marscher85, maraschi92, dermer92}
or alternatively with hadronic processes, for example, pion decay or proton synchrotron 
\citep{mannheim93a, muecke03,mannheim93b, aharonian2000}.

\begin{table*}
\caption{Model parameters for the one-zone SSC model SED of \pks0625 ($z=0.056$). The curve is depicted in Fig.~\ref{SED_0625}. Parameters marked with an asterisk were fixed during the fit.}         
\label{table:modelparameter}      
\centering                         
\begin{tabular}{ c c c c c c c c c c c c}        
\hline\hline
$\Gamma_\mathrm{j}^\ast$ & $\delta^\ast$ & $\theta^\ast$ & $p_1$ & $p_2$ & $E_\mathrm{min}$ & $E_\mathrm{max}$ & $E_\mathrm{break}$ & $U_e$ & $\eta$ & $B$ & $R^\ast$ \\
  &	    & [$^\circ$]&      &	&log10($E$/eV)	 & log10($E$/eV)	    & log10($E$/eV)        & [erg/cm$^{3}$] & [$U_e/U_B$]     & [G] & [cm] \\
\hline

 5	 & 6.7	    &  8       & 2.2   & 3.2	 & 7.4          & 11.7                   & 9.6               &  4.5 & 12.6  & $3.0$ &  $6.0\times10^{14}$      \\   

\hline

\end{tabular}
\end{table*}

We will apply a simple leptonic one-zone synchrotron-self Compton (SSC) model to the broad-band SED \citep{krawczynski04}. The parameters of the model are: the bulk Lorentz factor $\Gamma_\mathrm{j}$, the viewing angle $\theta$, the magnetic field $B$, the radius of the emission region $R$, the energy density of electrons $U_e$, the ratio $\eta$ of $U_e$ to the magnetic field energy density   $U_B$, the minimal and maximal electron energy $E_\mathrm{min}$ and $E_\mathrm{max}$, and the break energy $E_\mathrm{break}$. The electrons are assumed to follow a power-law energy spectrum of the form of $\mathrm{d}N/\mathrm{d}E\propto E^{-p_i}$ ($E$ electron energy in the jet frame)  with index $p_1$ for  $E_\mathrm{min}<E<E_\mathrm{break}$ and $p_2$ for $E_\mathrm{break}<E<E_\mathrm{max}$.  
The host galaxy emission dominates in the optical and part of the UV range.
Following \cite{PKS0625discovery} the host galaxy emission is included in the these energy ranges, by simulating the emission of a giant elliptical galaxy with the PEGASE 2 templates \citep{Fioc}.

For the SED modeling, we used only the data that were taken contemporaneously to the H.E.S.S. observations (blue points in Fig.\,\ref{SED_0625}), excluding the data from 2018-11-01 (red points in Fig.\,\ref{SED_0625}). As presented in this work, no significant multi-wavelength flux variations are found in the time range of 2018-11-02 to 2018-11-10 apart from the VHE flux changes. We assumed a small Lorentz factor of $\Gamma_\mathrm{j}=5$ and a small viewing angle of $\theta=8^\circ$ which results in a moderate boosting of $\delta=6.7$.  Furthermore, we used a radius of $R=6.0\times10^{14}$\,cm. Three of these assumed parameters, $\Gamma_\mathrm{j}$, $\delta$, and $R$, also marked with an asterisk in Table\,\ref{table:modelparameter}, result from variability constraints discussed in Sec.~3.1 whereas $\theta$ is derived from the Lorentz and Doppler factor constraints. We note that the value assumed for $R$ is comparable to radius estimates for the blazar Mrk\,421 at various time periods \citep{Mankuzhiyil2011}, but an order of magnitude smaller than for other TeV detected BL Lacertae-type objects \citep{Tavecchio2010, Mankuzhiyil2012} which typically showed VHE variability on longer time-scales as presented here for \pks0625. Also \cite{fukazawa15} found values of $R$ one or two orders of magnitude larger for PKS\,0625--354 for a five-year time period of  \textit{Fermi}-LAT data including a moderate flare in HE with a longer variability time scale. The absorption by the extragalactic background light (EBL) is taken into account in the SED with EBL-corrected VHE data points using the model by \citet{dominguez11}. The SED model presented here is only one possibility out of a large number of models suitable to explain the multi-wavelength data.
The parameters have been chosen to reproduce the shape of the lower energy component of the SED, especially the Swift-XRT data, and simultaneously to fit the high energy data points from Fermi-LAT and H.E.S.S. The SED as well as the model are shown in Fig.\,\ref{SED_0625}. The individual model parameters are given in Table\,\ref{table:modelparameter}. The resulting value for the $B$-field is rather high, but still consistent with values found by \cite{Mankuzhiyil2011}, \cite{Tavecchio2010}, and \cite{Ghisellini2011}. A comparison of the parameter results presented in this work for \pks0625 with SSC model parameter results from other misaligned objects is difficult because the sample of broad-band SEDs from misaligned objects based on simultaneous multiwavelength observations is very low. Moreover, comparing, for example, the results of $R$ for \pks0625 with SEDs of VHE detected BL Lacertae-type objects is difficult because sufficiently covered SEDs for BL Lacertae-type objects on a broad frequency range are typically based on time periods where the objects are not variable or variable on longer time periods in VHE.  

The SED presented here shows that the simple one-zone SSC model can fit the data reasonably well, similar to the results from \cite{fukazawa15}, where only contemporaneous X-ray and long-term Fermi-LAT data were considered for the modeling.
In contrast, the modeling by \cite{PKS0625discovery} indicated difficulties when using a simple SSC model based on contemporaneous data in the optical/UV, HE (long-term), and VHE data.
In this work we are able to consider nearly simultaneous optical/UV, X-ray, HE, and VHE data for the first time. 
 
In terms of the energy parameters of the electrons, $E_\mathrm{min}$ and $E_\mathrm{max}$, presented in this work, the model is similar to what is typically found for high-frequency peaked BL Lacertae-type objects (HBL) with a rather hard particle distribution, for example, \citet{hess1}, \cite{veritas13}, \cite{magic_1424},  \cite{magic15}, \cite{hess4}. Assuming that the break in the electron distribution is is the cooling break, resulting from an equilibrium between injection, radiative (synchrotron and inverse Compton) cooling, and escape, a spectral index of the injected electron distribution of 2.2 is required.
Furthermore, the model is Compton-dominated, which is commonly observed for low frequency peaked BL Lacertae-type objects and Flat Spectrum Radio Quasars {\citep{Fossati1998, Prandini2022}. 

Despite the limited statistics in the \textit{Fermi}-band - considering the break between the long-term \textit{Fermi} data and the spectrum obtained with H.E.S.S. - that flux level is comparable to the synchrotron peak. These findings show the very diverse characteristics of \pks0625 making a final classification of this object difficult. 

\begin{figure}
   \centering
   \includegraphics[width=9.cm]{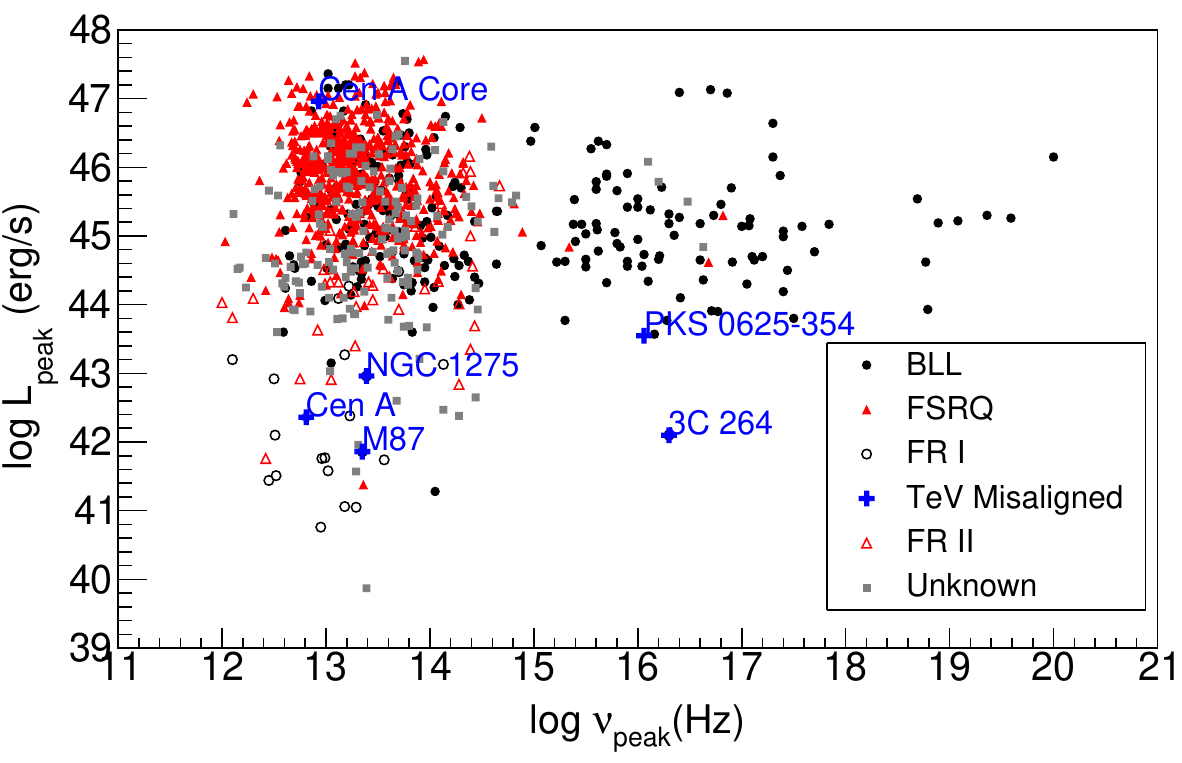}
   \caption{Synchrotron peak luminosity versus peak frequency as shown in \cite{Keenan2021}. BL Lacertae-type type objects are indicated with black filled circles while flat spectrum radio quasars are shown with red triangles. FR~I and FR~II radio galaxies are shown with black open circles and red open triangles, respectively. Objects of unknown type are plotted with gray squares. We highlight all TeV detected misaligned objects and included \pks0625 from this work (the object IC\,310 is not included in this figure due to an unknown peak frequency and peak luminosity). }
              \label{Meyer}
    \end{figure}

Previously, \cite{Keenan2021} presented a study of the synchrotron peak luminosity versus peak frequency plane and the dependence of the location in this plane on the jet power based on a large sample of AGNs. They discussed also the influence of velocity gradients of jets and the effect of misalignment and highlighted that all radio galaxies, except for 3C\,264 (independent of their power) appear at low synchrotron peak frequencies. In this work, we included \pks0625 (see Fig.~\ref{Meyer}) and further highlight TeV detected radio galaxies or misaligned blazars. The paucity of multiwavelength data from \pks0625 is likely the reason why this object was not included by  \cite{Keenan2021}. With the observation campaign in November 2018, presented in this paper the multiwavelength information is now improved, and thus we can estimate where \pks0625 is located in synchrotron the peak frequency - peak luminosity plane. Based on our broad-band spectral energy distribution we applied a simple log-parabolic fit to the Tanami data\footnote{We only include Tanami data in the radio band for the fit because the other radio measurements result from large radio beams that also integrate  the steep-spectrum diffusive emission from the extended jet.} from \cite{angioni19}, the high frequency UVOT data (which are less affected by the host galaxy emission) and XRT data, see Fig.~\ref{SED_0625}. The resulting peak frequency of $1.2\times10^{16}$\,Hz and its amplitude corresponding to $3.5\times10^{43}$\,erg\,s$^{-1}$ are shown in Fig.~\ref{Meyer}. Similar to 3C\,264, \pks0625 could represent a slightly less-beamed low-power jet with a significant velocity gradient along the jet axis \citep{Georganopoulos2003}  based on the location in the plane. The velocity gradient is consistent with  a low bulk Lorentz factor in order to obtain a rather large Doppler boosting in misaligned jets, which is needed to explain the fast TeV variability. The log-parabolic fit includes radio observations that are not simultaneous or contemporaneous to the UVOT and XRT data. Excluding the radio data in the fit results in a peak frequency of $1.9\times10^{15}$\,Hz and a synchrotron peak luminosity of $3.2\times10^{43}$\,erg\,s$^{-1}$, i.e., at about the same luminosity, but around one order of magnitude lower in frequency. This result does not affect the discussion and conclusions in this section. 

\section{Conclusions}

Simultaneous multi-wavelength observations and the investigation of the broad-band spectral energy distribution of non-blazar objects is an important tool to understand the $\gamma$-ray emission of AGNs in general. \pks0625 is one of only a few objects detected up to TeV energies, but no simultaneous optical/UV, X-ray, HE, and VHE observation campaign was performed before.

In this paper, we present multi-wavelength observations of \pks0625 conducted in November  2018. The light curve obtained with the H.E.S.S. telescopes revealed an outburst on November 1, 2018 followed by a decrease of the flux that can be described by an exponential decay with a flux halving time scale of $(5.9\pm2.7)$\,h. Unfortunately, there are no VHE observations covering the beginning of the outburst.
 
The VHE photon spectral index is consistent within the errors with the low-state spectrum and with the spectrum measured in 2012. The fast variability and the $\gamma\gamma$ pair production limits the viewing angle to $\sim10^\circ$, which is smaller than the upper limit obtained for this object from high-resolution radio measurements. 
Using \textit{Swift}-UVOT and -XRT data as well as Fermi-LAT measurements together with the H.E.S.S. results we reconstructed a simultaneous broad-band SED and fit this with a single-zone SSC model different to that presented in  \citet{PKS0625discovery}, where the X-ray and $\gamma$-ray data were non-simultaneous. 

The diverse multi-wavelength behavior of the object by showing one typical AGN-type characteristic in one frequency band and showing a different picture in another band - for example, extended radio emission versus a typical BL Lacertae-type type optical spectrum and variable gamma-ray emission - places \pks0625 at the borderline between different classes. It is likely that this is caused by a viewing angle that is too large to be a blazar and too small to be a typical radio galaxy. In principle, assuming the unified model of AGN, jet viewing angles should be randomly distributed in the universe, thus borderline objects must exist. Studying such individual objects is essential because they provide important constraints on acceleration and emission models such as the size and the location of the gamma-ray emission region. They often challenge the current understanding of the standard models for (high-energy) emission, for example, with flux variability, and thus they can help us to understand the physical processes in AGN jets in general.

\begin{acknowledgements}

The support of the Namibian authorities and of the University of
Namibia in facilitating the construction and operation of H.E.S.S.
is gratefully acknowledged, as is the support by the German
Ministry for Education and Research (BMBF), the Max Planck Society,
the Helmholtz Association, the French Ministry of
Higher Education, Research and Innovation, the Centre National de
la Recherche Scientifique (CNRS/IN2P3 and CNRS/INSU), the
Commissariat à l’énergie atomique et aux énergies alternatives
(CEA), the U.K. Science and Technology Facilities Council (STFC),
the Irish Research Council (IRC) and the Science Foundation Ireland
(SFI), the Polish Ministry of Education and Science, agreement no.
2021/WK/06, the South African Department of Science and Innovation and
National Research Foundation, the University of Namibia, the National
Commission on Research, Science \& Technology of Namibia (NCRST),
the Austrian Federal Ministry of Education, Science and Research
and the Austrian Science Fund (FWF), the Australian Research
Council (ARC), the Japan Society for the Promotion of Science, the
University of Amsterdam and the Science Committee of Armenia grant
21AG-1C085. We appreciate the excellent work of the technical
support staff in Berlin, Zeuthen, Heidelberg, Palaiseau, Paris,
Saclay, Tübingen and in Namibia in the construction and operation
of the equipment. This work benefited from services provided by the
H.E.S.S. Virtual Organisation, supported by the national resource
providers of the EGI Federation.\\

This publication makes use of data products from the Two Micron All Sky Survey, which is a joint project of the University of Massachusetts and the Infrared Processing and Analysis Center/California Institute of Technology, funded by the National Aeronautics and Space Administration and the National Science Foundation.

\end{acknowledgements}

\bibliographystyle{aa} 
\bibliography{references.bib}

\end{document}